\documentclass{svjour2}                    
\smartqed  
\usepackage{graphicx}
\def\bx{{\bf x}}

\begin{document}

\title{Frequency estimate for multicomponent crystalline compounds.
\thanks{Supported by the Department of Energy under grant DE-SC0014506}
}


\author{M. Widom}


\institute{M. Widom \at
              Department of Physics, Carnegie Mellon University, 
              Pittsburgh, PA  15213 
}

\date{Received: date / Accepted: date}

\maketitle

\begin{abstract}
Among crystal structures of $N$-component metal alloys, far fewer examples are known with $N\ge 4$ than with $N=2$ or 3, in apparent contradiction to the exponentially growing number of possible combinations of elements.  Two effects contribute to this shortfall.  Since the $N$-component composition space resides within a $d$-dimensional simplex with $d=N-1$, the vanishing volume in high dimensions reduces the distinct $N$-component compositions.  Additionally, the increasing surface area makes it more probable that stable structures reside on the surface of the simplex (containing fewer than $N$ components) as opposed to its interior.  Specific estimates are developed through application of the empirical Miedema enthalpy model.  Despite their rarity, we propose that the actual number of $N=4$- and 5-component alloys greatly exceeds the number that are currently known.

\end{abstract}

\section{Introduction}
\label{intro}

\begin{figure}
\includegraphics[angle=-90,width=4in]{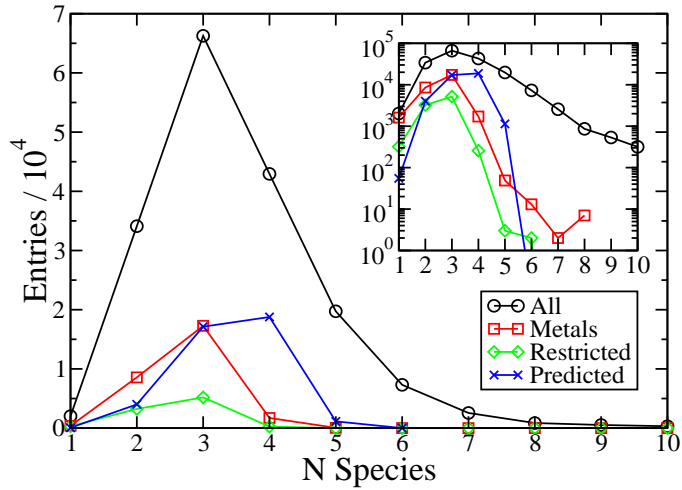}
\caption{Distribution of ICSD entries with respect to the number of elements (data for year 2016). "All" includes $E=105$ elements, while ``Metals'' excludes nonmetals H, B, C, N, P, chalcogenides, halides and noble gases resulting in $E=84$. ``Restricted'' further restricts to the $E=55$ metallic elements with Miedema parameters~\cite{Miedema1980} and eliminates solid solutions. ``Predicted'' results from our model as discussed below. Numerical values for restricted and predicted curves are presented in Table~\ref{tab:results}.}
\label{fig:icsd}
\end{figure}

Many thousands of inorganic crystal structures are currently known. The Inorganic Crystal Structure Database (ICSD), for example, lists more than 180,000 entries and adds approximately 6,000 each year.  Consider the distribution of entries with respect to the number of elements, $N$, as illustrated in Fig~\ref{fig:icsd}.  Given $E$ elements of the periodic table, there are $N_s={E \choose N}$ combinations of elements each of which defines an $N$-component (or ``$N$-ary'') alloy system.  For $N\ll E$ this number grows exponentially as $E^N$. Further, for each $N$-component alloy system, many specific compositions (relative concentrations of elements) are possible, so that $N_s$ could greatly underestimate the number of distinct compounds.  Strikingly, the distributions in Fig.~\ref{fig:icsd} reveal vanishing numbers of multicomponent crystals, in contrast to the expected exponential growth.

The composition of an $N$-component compound can be represented by a vector $\bx=(x_1,x_2,\dots,x_N)$, where $x_\alpha\ge 0$ is the mole fraction of chemical species $\alpha$, and hence $\sum_\alpha x_\alpha = 1$.  For a given $N$, the set of all possible $\bx$ is a $d$-dimensional regular simplex of unit edge length with $d=N-1$.  The set of stable compounds is given by the subset of this simplex that lies on the convex hull of free energy.  We shall focus on structures that are stable in the limit of low temperature, so the free energy is simply the enthalpy, and the convex hull is expected to contain a set of discrete vertices corresponding to the stable crystal structures~\cite{Mihal04}.

In this paper we identify two geometrical attributes of the $d$-dimensional simplex that contribute to the shortfall in $N$-ary crystal structures.  First, genuine $N$-component compounds, requiring that every species be present in some fraction $x_\alpha>0$, reside in the $d$-dimensional {\em interior} of the simplex. The volume of the $d$-dimensional unit simplex is $V_d=\sqrt{d+1}/d!\sqrt{2^d}$, leading to a faster than exponential decrease of the available composition space.  Second, the $d'$-dimensional surface of the unit simplex (where one or more $x_\alpha$ vanishes, so that $d'<d$) greatly exceeds its $d$-dimensional volume. For example there are $d(d+1)/2$ edges, each of length $V_{d'=1}=1$, along which only two mole fractions are nonzero.  The rapid growth surface to volume ratio for large $N$ makes it likely that vertices of the convex hull lie on the surface of the simplex corresponding to an $L$-ary compound with $L<N$ rather than a true $N$-ary.

We test these conjectures by evaluating the enthalpies of hypothetical $N$-ary crystal structures using the empirical Miedema enthalpy model~\cite{Miedema1980}.  Our predicted frequency of occurrence of stable $N$-ary compounds vanishes for large $N$ in accordance with the above observations.  However, for $N=4$ and 5, we propose that the actual numbers of stable $N$-ary compounds, despite their scarcity, greatly exceed the numbers reported in the ICSD.  Hence many interesting and potentially useful new compounds might remain to be discovered.

\section{Model}
\label{sec:model}

We seek to count the number of crystal structures stable at low temperatures formed by combinations of metallic elements.  For the sake of definiteness, we assume these are stoichiometric compounds of unique fixed composition and vanishing entropy.  Specifically, at most one such structure exists at any given composition, establishing a 1:1 correspondence between stable structures and stable compositions.

We do not know in advance what these structures are since in many cases they have yet to be discovered.  Some automated methods exist to discover actual structures, provided that an energy model exists~\cite{Glass06,Fredeman11,vanDeWalle02_1,Woodley2008}. Considering the exponentially large number $N_s$ of alloy systems we will investigate, developing interatomic potentials is impractical and application of first principles methods is prohibitively time consuming. Rather, we abandon the idea of identifying specific crystal structures and focus only on the problem of counting stable compositions. Additionally, we will not insist on an exact count, but rather make an estimate. This will be achieved by discretizing the composition space, then applying a preexisting universal, structure agnostic, model for the enthalpy on this space.

\subsection{Discrete composition space}
\label{sec:discrete}

For a given alloy system, the set of stable crystal structures implies a discrete set of stable compositions $\{\bx_i\}$ with $N_c$ members where $i=1,\dots,N_c$ indexing the different compounds. If there are finitely many stable compositions there must be minimum separations among the $\bx_i$\footnote{Certain artificial models exhibit infinitely many stable compositions~\cite{Bak1982} but no clear case of such behavior has been found experimentally.}.  Every alloy system has its own set $\{\bx_i\}$.  For a given number of species $N$ there exists an average value of $N_c$, and an average minimum separation of $\{\bx_i\}$.  We impose this by requiring that every mole fraction be a rational fraction\footnote{Hence we exclude quasicrystals~\cite{Shechtman1984,Levine1984} which are few in number and might not be stable at low temperatures~\cite{Widom1991,deBoissieu1995}.} with a common denominator, i.e. $x_\alpha=m_\alpha/M$ with $\sum_\alpha m_\alpha=M$. We place every set $\{m_\alpha\}$ in unique correspondence with one potentially stable crystal structure.

Discretization limits the number of $N$-ary compounds because, if $N>M$, no composition exists with $N$ nonzero integers $m_\alpha$ summing to $M$.  More generally, we can enumerate the possible $N$-ary compositions.  The first case is binary compounds with $N=2$ in a composition space of dimension $d=N-1=1$.  There are $M+1$ possible values of $m_1$ spaced uniformly on the interval from 0 to $M$, while $m_2=M-m_1$ is uniquely determined. However, of these $M+1$ discrete compositions, two of them ($m_1=0,M$) lie on the boundary of the interval and thus correspond to pure elements, leaving $M-1$ interior points corresponding to true binaries.

The next case is ternary compounds with $N=3$ in a composition space of dimension $d=N-1=2$.  Possible values of $m_1$ again range from 0 to $M$, values of $m_2$ range from 0 to $M-m_1$, while values of $m_3=M-m_1-m_2$ are uniquely determined. Their total numbers of discrete compositions are given by the triangle number ${\rm Tr}_{M+1}$, where ${\rm Tr}_n\equiv{n+1\choose 2}$.  Dropping the boundary cases corresponding to triangle edges and vertices results in a smaller triangle; the number of interior points is ${\rm Tr}_{M-2}$.  Similarly for quaternaries with $N=4$ the composition space has dimension $d=N-1=3$, the numbers of discrete compositions are given by tetrahedral numbers ${\rm Te}_{M+1}$ where ${\rm Te}_n={n+2\choose 3}$, while dropping tetrahedron faces, edges and vertices results in a number of interior points ${\rm Te}_{M-3}$.

For arbitrary $M$ and $N$, the total number of points is ${N+M-1 \choose N-1}$ and the number of interior points is ${M-1 \choose N-1}$.  The remaining boundary points correspond to $L$-ary alloy subsystems with $L\ge N-1$.  At fixed denominator $M$, the number of interior points, which represents the maximum number of possible stable $N$-ary compounds in our model, grows linearly from the value 1 at $N=1$, reaches a maximum at $N=1+M/2$, then vanishes at $N=M+1$.  Thus, despite the faster than exponential decrease of interior volume of the unit simplex, $V_d$, the number of interior points {\em increases} with $N$ as $M^{N-1}$ for $N\ll M/2$ owing to the density of discrete compositions with denominator $M$.

To choose a value for $M$, we examined values of $\sum_\alpha m_\alpha$ among the stoichiometric structures in the ICSD, defined as those with only integral values of $m_\alpha$.  The top twelve most common metal structure types exhibit $M\le 6$, but the distributions are broad, with median values below $M=10$ but mean values above.  Both medians and means grow slightly with increasing numbers of constituent elements $N$.  Here, for simplicity, we take $M=10$ as a common value for all $N$-ary alloy systems.

\subsection{Miedema enthalpy model}
\label{sec:miedema}

Low temperature stable crystal structures minimize the enthalpy $H$.  If we could calculate the enthalpy for all possible crystal structures within a given $N$-ary alloy system, the stable structures would occupy the vertices of the convex hull of the enthalpy~\cite{Mihal04}.  In the present case we do not know the precise structures whose enthalpy is required.  Even if we knew the structures, they are too numerous to carry out first principles total energy calculations, and the alloy systems are too diverse for reliable application of fitted interatomic potentials~\cite{Ercolessi1994,Mihal2012}.

Thus we adopt the empirical Miedema model~\cite{Miedema1980}, which approximates the enthalpy without specification of the crystal structure.  Pairwise interatomic interactions are expressed as functions of the atomic volume, electronegativity, electron density and other properties that have been tabulated for $E=55$ individual metallic elements.  For a binary alloy of species $\alpha$ and $\beta$, the chemical enthalpy of mixing can be expressed~\cite{Mousavi2016} as
\begin{equation}
\label{eq:Hchem}
\Delta H^{\rm Chem}_{\alpha\beta} = f(x_\alpha,x_\beta)\left(x_\alpha\Delta H^{\rm Sol}_{\alpha~{\rm in}~\beta}
+x_\beta\Delta H^{\rm Sol}_{\beta~{\rm in}~\alpha}\right),
\end{equation}
where $\Delta H^{\rm Sol}{\alpha~{\rm in}~\beta}$ is the enthalpy of solution, and
\begin{equation}
f(x^S_\alpha,x^S_\beta)=x^S_\alpha x^S_\beta
\left[1+\delta(x^S_\alpha x^S_\beta)\right]
\end{equation}
is a function of the surface concentrations $x^S_\alpha = x_\alpha V_\alpha^{2/3}/\sum_\beta x_\beta V_\beta^{2/3}$.  Here $V_\alpha$ is the molar volume of species $\alpha$, and the parameter $\delta$ is a structure-dependent term taking values 0 for solid solutions, 5 for amorphous structures and 8 for ordered structures. Large values of $\delta$ make $\Delta H^{\rm Chem}$ nonconvex as a function of $x_\alpha$.  In ignorance of the specific structure, we will take $\delta$ as a random value from 0 to 8 assigned independently to every potential structure.

Although originally developed and tested for binary alloys, $\Delta H^{\rm Chem}$ can be naturally generalized to multicomponent systems~\cite{Goncalves1996,Wang2009,Sadeghi2013,Mousavi2016} as
\begin{equation}
\label{eq:Hchemmulti}
\Delta H^{\rm Chem} = \sum_{\alpha\ne \beta=1}^N x_\alpha f(x_\alpha,x_\beta) H^{\rm Sol}_{\alpha~{\rm in}~\beta}.
\end{equation}
The range of $\delta$ values in $f(x^S_\alpha,x^S_\beta)$ must be scaled by $(N/2)^2$ to maintain its strength as $N$ grows, owing to the $1/N$ fall-off of the surface concentrations $X^S_\alpha$.

The Miedema model formation enthalpy contains two additional terms, $\Delta H^{\rm Elastic}$ for elastic strain of solid solutions~\cite{Niessen1983}, and $\Delta H^{\rm Struct}$ that discriminates between FCC, HCP and BCC structures of transition metals and depends on their valence electron counts~\cite{Miedema1983}. We omit the elastic term because we are not interested in solid solutions.  Further, because we do not know the crystal structure, we randomly assign one of the three choices, and weight the contribution by the mole fraction of each transition metal.  Actual multicomponent structures are frequently complex crystal types with many inequivalent Wyckoff positions~\cite{Steurer2016}, hence this term should be considered as a random variable whose magnitude is typical of actual transition metal interactions.  For the pure elements whose structures {\em are} known, we utilize their proper structural energies.

\section{Results}
\label{sec:results}

We consider all $N_s$ alloy systems containing $N=1-6$ elements for the $E=55$ elements with Miedema parameters. In every case we calculate the enthalpies of all ${N+M-1\choose N-1}$ discrete compositions that include ${M-1\choose N-1}$ true $N$-ary compositions in addition to their $L$-ary alloy subsystems with $L<N$.  The convex hull~\cite{qhull} of this set of enthalpies over the composition simplex in general contains many vertices corresponding to stable compositions.

\begin{table}
\caption{Statistical summary of results. Column $N_s={E\choose N}$ is the number of $N$-ary alloy systems formed among $E=55$ metallic elements. Column $N$-aries tabulates ${E\choose N}{M-1\choose N-1}$, the number of true $N$-ary combinations of integers such that $\sum_\alpha m_\alpha=M$ with all $m_\alpha>0$ and $M=10$. Column $V/S$ is the volume/surface fraction of potential compositions.  Predicted is the number of predicted stable $N$-aries on the convex hull of Miedema enthalpies. ICSD is the number of unique structures listed in the ICSD containing metallic elements with Miedema parameters.}
\label{tab:results}       
\begin{tabular}{l|rrr|rr}
\hline\noalign{\smallskip}
$N$  & $N_s$         & $N$-aries   & $V/S$  & Predicted & ICSD \\
\noalign{\smallskip}\hline\noalign{\smallskip}
1    & 55            & 55          & NA     & 55        & 321  \\
2    & 1485          & 13365       &4.50    & 4009      & 3236 \\
3    & 26235         & 944460      &1.20    & 17126     & 5187 \\
4    & 341055        & 28648620    &0.42    & 18775     & 257  \\
5    & 3478761       & 438323886   &0.14    & 1137      & 3    \\
6    & 28989675      & 3652699050  &0.04    & 0         & 2    \\
\noalign{\smallskip}\hline
\end{tabular}
\end{table}

\begin{figure}
\includegraphics[angle=-90,width=4in]{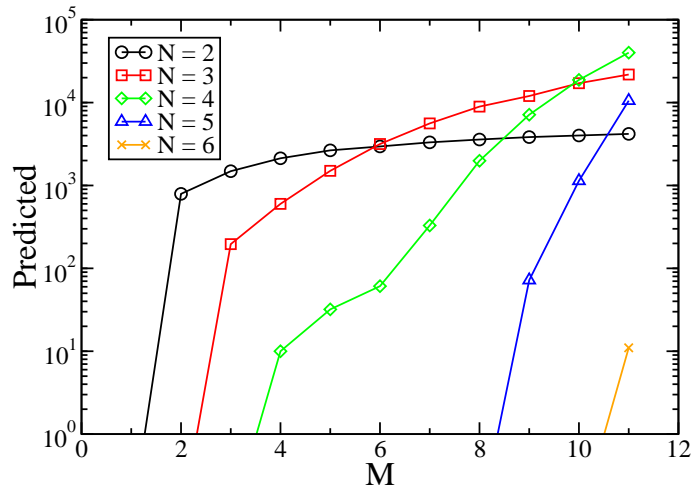}
\caption{Dependence of predicted stable $N$-aries on the discretization denominator $M$.}
\label{fig:NvM}
\end{figure}

Table~\ref{tab:results} summarizes our statistical findings.  Notice that, despite the exponential growth of the number of alloy systems and potential $N$-ary structures, the number of predicted stable structures reaches a maximum at $N=4$ and essentially vanishes for $N\ge 6$.  In comparison, the corresponding numbers listed in the ICSD peak at $N=3$ and fall off more rapidly than our estimate.  Of course our estimates depend on the value of $M$ that we set in an ad-hoc manner.  The predicted number of binaries varies only slightly with $M$, while the number of quinaries and hexaries, for example, vary dramatically as illustrated in Fig.~\ref{fig:NvM}.

The increasing number of multicomponent alloys for $N=2-4$ results from the two order of magnitude increase in the number of potential alloy systems together with an additional order of magnitude increase in the number of distinct compositions.  More interesting is the {\em decrease} in predicted stable structures beyond $N=4$.  This is due in part to the decreasing volume fraction of interior points in the $(N-1)$-dimensional composition space.  If only a small fraction of compositions are true $N$-aries, then a correspondingly small fraction are likely to lie on the convex hull.

\begin{figure}
\includegraphics[angle=-90,width=4in]{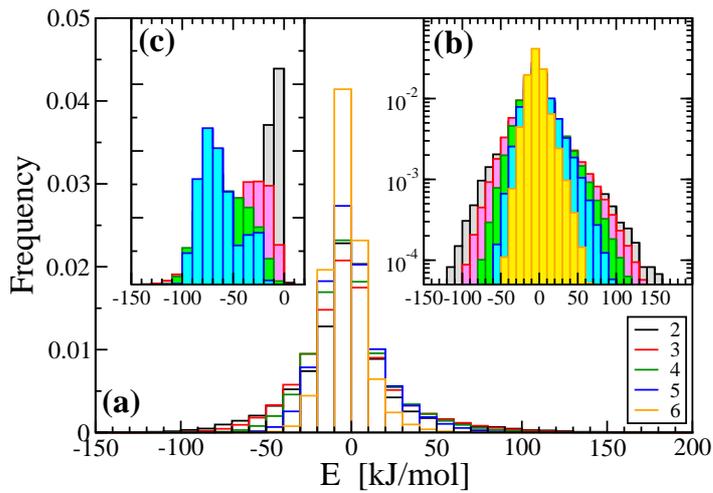}
\caption{Normalized distributions of predicted formation enthalpies for multicomponent alloys. (a) Complete distributions on a linear scale; (b) Complete distributions on a logarithmic scale; (c) Distributions for stable compounds on a linear scale. Note 100 kJ/mol = 1.04 eV/atom.}
\label{fig:distrib}
\end{figure}

Table~\ref{tab:results} shows that the actual shortfall of stable $N$-aries is far more extreme than the volume fraction would suggest; a second effect is present.  As shown in Fig.~\ref{fig:distrib}a and b, the predicted formation enthalpies belong to distributions whose width narrows as $N$ increases.  Table~\ref{tab:distrib} lists the means and standard deviations of these distributions.  The means nearly vanish and the standard distributions fall off approximately as $1/\sqrt{N}$ for large $N$, as expected from the form of Eq.~(\ref{eq:Hchemmulti}) that represents the multicomponent enthalpy as a weighted average of binary enthalpies.  Binary compounds are more likely to be highly stable than ternaries, for example.  Indeed, in an $N=6$ component alloy system we predict four times more stable binary subsystems than ternaries, and an order of magnitude more ternaries than quaternaries, and quaternaries than quinaries.

\begin{table}
\caption{Statistics of enthalpy distributions (kJ/mol).}
\label{tab:distrib}       
\begin{tabular}{l|rr|rr}
\hline\noalign{\smallskip}
     & \multicolumn{2}{c|}{All} & \multicolumn{2}{c}{Stable} \\
$N$  & Mean & Deviation        & Mean & Deviation \\
\noalign{\smallskip}\hline\noalign{\smallskip}
2    & -2.61 & 31.1    & -23.4 & 21.8 \\
3    & -1.86 & 28.7    & -43.1 & 25.4 \\
4    & -0.85 & 23.9    & -57.6 & 22.2 \\
5    & -0.08 & 19.2    & -63.8 & 19.1 \\
6    & -1.09 & 13.4    & NA    & NA   \\
\noalign{\smallskip}\hline
\end{tabular}
\end{table}

Within our model, stability of a many-component alloy requires that random fluctuations conspire to push the enthalpy to the extreme lower limit of its distribution in order to beat the enthalpies of competing fewer-component alloy subsystems whose enthalpies are drawn from broader ranges.  This is illustrated in Fig.~\ref{fig:distrib}c, where the distributions for stable compositions (those that reach the convex hull) show that the average enthalpies of stable multicomponent structures lie in the low enthalpy tails of the distributions of all enthalpies.  The tails of these distributions, together with correlation of the many-component alloy enthalpy with the enthalpies of its competing few-component alloy subsystems, reduce the number of predicted stable compositions as shown by Table~\ref{tab:results}.

\section{Discussion}
\label{sec:discussion}

We have identified two geometric properties of multicomponent composition spaces that limit the number of low temperature stable $N$-ary compounds for large $N$: the shrinking volume, and the diverging surface area of the $d$-dimensional simplex representing the composition space.  These principles were illustrated by estimating the frequency of stable compositions within a model that discretized the composition space and applied the empirical Miedema enthalpy model.  The chief finding is that, for large $N$, stable compounds (vertices of the convex hull of enthalpies) overwhelmingly lie on the surface of the simplex rather than its interior.

The numbers of known quaternary and quinary crystal structures fall dramatically short of our estimates.  While we can vary our predictions by adjusting the value of $M$, the case for {\em increasing} $M$ is stronger than the case for decreasing it, as discussed at the end of Section~\ref{sec:discrete}.  Likely the discrepancy is real, and many stable metallic crystals remain to be discovered.  Keep in mind, also, that we excluded the majority of rare-earth elements from our study.  With new discoveries of interesting physical properties regularly reported, even among compounds with few elements, it is quite possible that a large number of practically useful, commercially important multicomponent crystal structures remain to be discovered~\cite{DiSalvo2000}.

Several factors may contribute to the small numbers of known many-component crystal structures.  First, experimental effort has been weighted towards binary and ternary compounds.  Many-component studies are often restricted to the {\em boundaries} of composition space, with most constituents present only in trace quantities.  For example, steels contain primarily Fe with a few weight-percent C plus minor additions of Mn, Ni, Cr, etc. Likewise, Nickel-based superalloys are primarily Ni$_3$Al with minor addition of Cr, Fe, Co, Mo, etc.

The discovery of high entropy alloys (HEAs) containing many principle elements~\cite{Cantor2004,Yeh2004} sparked intensive investigation of the {\em interior} of multicomponent composition spaces.  HEAs invoke substitutional configurational entropy of $\log{(N)}$ at high temperature to stabilize multicomponent solid solutions against phase separation into competing crystalline structures with fewer elements. HEAs are thus not generally expected to form low temperature stable structures.  One exception is Mo-Nb-Ta-W~\cite{Senkov10}, where previously unknown quaternary ground states were found~\cite{Widom2016}.

A second factor limiting the discovery of multicomponent crystal structures lies in the complex character of alloy phase diagrams.  Numerous phases can exist over narrow composition ranges, such as in e.g. Al-Co or Cd-Yb, requiring fine tuning of composition and temperature to form certain phases from the melt. Additionally phases may exist only at low temperatures and never coexist with the melt, as for example in Al$_5$Ti$_3$ and Al$_4$Zr$_3$.  In such cases the structure can only be discovered by careful annealing, which is time consuming and unlikely to be performed unless a specific new phase is suspected.  Large $N$ crystalline ground states are likely to be complex structures that are difficult to form by conventional methods, potentially limiting their practical application.

Finally, even if serious efforts were made to investigate multicomponent alloy phase diagrams, the sheer number of such systems (see column $N_s$ in Table~\ref{tab:results}) shows that it will be difficult to sample all combinations.

Some computational strategies can accelerate the search.  The CALPHAD method utilizes thermodynamic databases based on binary and ternary data, augmented with interpolation formulas for solid solutions of known crystal types, and hence is well suited to the discovery of new high entropy alloys stable at elevated temperatures.  Self-consistent databases for multi-principle component alloys are now being developed~\cite{Zhang2012,TCHEA1}.  However, the logarithmic increase of entropy with $N$ is insufficient to counter the extreme enthalpies of small $N$ compounds, so that only about 1.2\% of $N=6$ compounds are expected to form single phase solid solutions~\cite{Senkov2015}.

Alternative methods are required to discover novel structures.  We already mentioned high throughput computation using cluster expansions~\cite{vanDeWalle02_1} to speed the energy calculation, and genetic algorithms~\cite{Glass06,Fredeman11} to search configuration space.  An alternate strategy is to generalize based on known binary and ternary structures~\cite{Mihal04}. Every crystal structure consists of an underlying lattice decorated with atoms at symmetric positions belonging to Wyckoff classes~\cite{Steurer2016}.  Many structures possess multiple inequivalent Wyckoff classes, even though the same atomic species may occupy multiple Wyckoff classes.  In such a case it is possible that some other chemically similar element might readily occupy the same Wyckoff class.  A typical example is the prototype Al$_4$Ba with three Wyckoff classes two of which are occupied by Al.  This structure is intrinsically ternary, and is represented by the widely studied pnictide superconductor BaFe$_2$As$_2$.  The stability of this new multicomponent structure can easily be evaluated using density functional theory and then tested experimentally when warranted.

\begin{acknowledgements}
I thank Alan Frieze and Wesley Pegden for a discussion on generalized triangle numbers.  This paper is dedicated to the memory of Prof. Leo P. Kadanoff, my PhD thesis advisor.
\end{acknowledgements}

\bibliographystyle{./spmpsci}      
\bibliography{multi}   

\begin{thebibliography}{10}
\providecommand{\url}[1]{{#1}}
\providecommand{\urlprefix}{URL }
\expandafter\ifx\csname urlstyle\endcsname\relax
  \providecommand{\doi}[1]{DOI~\discretionary{}{}{}#1}\else
  \providecommand{\doi}{DOI~\discretionary{}{}{}\begingroup
  \urlstyle{rm}\Url}\fi

\bibitem{Bak1982}
Bak, P.: Commensurate phases, incommensurate phases and the devil's staircase.
\newblock Rep. Prog. Phys. \textbf{45}, 587--629 (1982)

\bibitem{qhull}
Barber, C.B., Dobkin, D.P., Huhdanpaa, H.T.: The quickhull algorithm for convex
  hulls.
\newblock ACM Trans. Math. Software \textbf{22}, 469--83 (1996)

\bibitem{deBoissieu1995}
de~Boissieu, M., Boudard, M., Hennion, B., Bellissent, R., Kycia, S., Goldman,
  A., Janot, C., Audier, M.: Diffuse scattering and phason elasticity in the
  alpdmn icosahedral phase.
\newblock Phys. Rev. Lett. \textbf{75}, 89--92 (1995)

\bibitem{Cantor2004}
Cantor, B., Chang, I.T.H., Knight, P., Vincent, A.J.B.: Microstructural
  development in equiatomic multicomponent alloys.
\newblock Mat. Sci. Eng. A  (2004)

\bibitem{DiSalvo2000}
DiSalvo, F.J.: Challenges and opportunities in solid-state chemistry.
\newblock Pure Appl. Chem. \textbf{72}, 1799--1807 (2000)

\bibitem{Ercolessi1994}
Ercolessi, F., Adams, J.B.: The force-matching method.
\newblock Europhys. Lett. \textbf{26}, 583 (1994)

\bibitem{Fredeman11}
Fredeman, D.J., Tobash, P.H., Torrez, M.A., Thompson, J.D., Bauer, E.D.,
  Ronning, F., Tipton, W., Rudin, S.P., , Hennig, R.G.: Computationally-driven
  experimental discovery of the {C}e{I}r4{I}n compound.
\newblock Phys. Rev. B \textbf{83}, 224,102 (2011)

\bibitem{Glass06}
Glass, C.W., Oganov, A.R., Hansen, N.: Uspex -- evolutionary crystal structure
  prediction.
\newblock Comp. Phys. Comm. \textbf{175}, 713--20 (2006)

\bibitem{Goncalves1996}
Goncalves, A.P., Almeida, M.: Extended {M}iedema model: Predicting the
  formation enthalpies of intermetallic phases with more than two elements.
\newblock Physica B \textbf{228}, 289--94 (1996)

\bibitem{Levine1984}
Levine, D., Steinhardt, P.J.: Quasicrystals: A new class of ordered structures.
\newblock Phys. Rev. Lett. \textbf{53}, 2477--80 (1984)

\bibitem{Miedema1980}
Miedema, A.R., de~Chatel, P.F., de~Boer, F.R.: Cohesion in alloys --
  fundamentals of a semi-empirical model.
\newblock Physica B \textbf{100}, 1--28 (1980)

\bibitem{Miedema1983}
Miedema, A.R., Niessen, A.K.: The enthalpy of solution for solid binary alloys
  of two 4d-transition metals.
\newblock CALPHAD \textbf{7}, 27--36 (1983)

\bibitem{Mihal2012}
Mihalkovic, M., Henley, C.L.: Empirical oscillating potentials for alloys from
  ab initio fits and the prediction of quasicrystal-related structures in the
  al-cu-sc system.
\newblock Phys. Rev. B \textbf{85}, 092,102 (2012)

\bibitem{Mihal04}
Mihalkovi\v{c}, M., Widom, M.: Ab-initio cohesive energies of {F}e-based
  glass-forming alloys.
\newblock Phys. Rev. B \textbf{70}, 144,107 (2004)

\bibitem{Mousavi2016}
Mousavi, M.S., Abbasi, R., Kashani-Bozorg, S.F.: A thermodynamic approach to
  predict formation enthalpies of ternary systems based on {M}iedema's model.
\newblock Met. Mat. Trans. A \textbf{47}, 3761--3770 (2016)

\bibitem{Niessen1983}
Niessen, A.K., Miedema, A.R.: The enthalpy effect on forming diluted solid
  solutions of two 4d and 5d transition metals.
\newblock Ber. Bunsenges. Phys. Chem. \textbf{87}, 717--25 (1983)

\bibitem{Sadeghi2013}
Sadeghi, E., Karimzadeh, F., Abbasi, M.H.: Thermodynamic analysis of
  {T}i{A}l{C} intermetallics formation by mechanical alloying.
\newblock J. Alloy Comp. \textbf{576}, 317--23 (2013)

\bibitem{Senkov2015}
Senkov, O.N., Miller, J.D., Miracle, D.B., Woodward, C.: Accelerated
  exploration of multi-principal element alloys with solid solution phases.
\newblock Nat. Comm. \textbf{6}, 6529 (2015)

\bibitem{Senkov10}
Senkova, O., Wilks, G., Miracle, D., Chuang, C., Liaw, P.: Refractory
  high-entropy alloys.
\newblock Intermetallics \textbf{18}, 1758--65 (2010)

\bibitem{Shechtman1984}
Shechtman, D., Blech, I., Gratias, D., Cahn, J.W.: Metallic phase with
  long-range orientational order and no translational symmetry.
\newblock Phys. Rev. Lett. \textbf{53}, 1951--3 (1984)

\bibitem{Steurer2016}
Steurer, W., Dshemuchadse, J.: Intermetallics: Structures, Properties, and
  Statistics.
\newblock Int. Union of Cryst. (2016)

\bibitem{TCHEA1}
ThermoCalc: {TCHEA1 - TCS High entropy alloy database, Version 1.0} (2015).
\newblock
  \urlprefix\url{http://www.thermocalc.com/media/35873/tchea10\_extended\_info\_bh.pdf}

\bibitem{vanDeWalle02_1}
van~de Walle, A., Ceder, G.: Automating first-principles phase diagram
  calculations.
\newblock J. Phase Equilib. \textbf{23}(4), 348--359 (2002)

\bibitem{Wang2009}
Wang, T.L., Liu, B.X.: Glass forming ability of the {F}e{Z}r{C}u system studied
  by thermodynamic calculation and ion beam mixing.
\newblock J. Alloy Comp. \textbf{481}, 156--60 (2009)

\bibitem{Widom1991}
Widom, M.: Elastic stability and diffuse scattering in icosahedral
  quasicrystals.
\newblock Phil. Mag. Lett. \textbf{64}, 297--305 (1991)

\bibitem{Widom2016}
Widom, M.: High Entropy Alloys: Fundamentals and applications, chap. 8.
  Prediction of structure and phase transformations, pp. 267--298.
\newblock Springer (2016).
\newblock Eds. Michael C. Gao, Jien-Wei Yeh, Peter K. Liaw and Yong Zhang

\bibitem{Woodley2008}
Woodley, S.M., Catlow, R.: Crystal structure prediction from first principles.
\newblock Nat. Mat. \textbf{7}, 937--46 (2008)

\bibitem{Yeh2004}
Yeh, J.W., Chen, S.K., Lin, S.J., Gan, J.Y., Chin, T.S., Shun, T.T., Tsau,
  C.H., Chang, S.Y.: Nanostructured high-entropy alloys with multiple principal
  elements: Novel alloy design concepts and outcomes.
\newblock Adv. Eng. Mater. \textbf{6}, 299--303 (2004)

\bibitem{Zhang2012}
Zhang, C., Zhang, F., Chen, S., Cao, W.: Computational thermodynamics aided
  high-entropy alloy design.
\newblock JOM \textbf{64}, 839--45 (2012)

\end{thebibliography}

\end{document}